\documentstyle[eqsecnum,aps]{revtex}
\begin{document}
\draft
\baselineskip= 12pt
\title{\bf
Nambu-Goldstone mechanism in real-time thermal field theory
\\
}
\author{
{\bf
Bang-Rong Zhou}
}
\address{  
Department of Physics, Graduate School at Beijing,
University of Science and Technology of China \\
Academia Sinica, Beijing 100039, China\thanks{Permanent address.} \\
and \\
The Abdus Salam International Centre for Theoretical Physics,
P.O.Box 586, 34100 Trieste, Italy \\
}
\date{}
\maketitle 
\begin{abstract}
In a one-generation fermion condensate scheme of electroweak symmetry breaking,
it is proved based on the Schwinger-Dyson equation in real-time thermal field
theory  in the fermion bubble diagram approximation  that,
at finite temperature $T$ below the symmetry restoration
temperature $T_c$, a massive Higgs boson and three massless Nambu-Goldstone
bosons could emerge from spontaneous breaking of the electroweak group
${\rm SU}_L(2)\times {\rm U}_Y(1) \to {\rm U}_Q(1)$ if the two fermion flavors
in the one
generation are mass degenerate, thus the Goldstone theorem is rigorously valid
in this case. However, if the two fermion flavors have unequal masses, owing to
"thermal fluctuation", the Goldstone theorem will be true only approximately
for a very large momentum cutoff $\Lambda$ in the zero temperature fermion loop
or for low energy scales. All possible pinch singularities are proved to
cancel each other, as expected in a real-time thermal field theory.
\end{abstract}

\section{Introduction}
Spontaneous symmetry breaking at finite temperature has been investigated
extensively [1-6].  However, most research by now has concentrated on the
discussions of phase transition and critical temperature based on the effective
potential approach at finite temperature, and relatively less work is
reported about the theoretical exploration of the Nambu-Goldstone theorem [7] at
finite temperature, especially in models of dynamical symmetry breaking
such as the Nambu-Jona-Lasinio (NJL) model with four-fermion interactions [7].
Research into the Nambu-Goldstone mechanism at finite temperature could provide
us a deeper understanding of the spontaneous breaking of a continuous symmetry at
finite temperature and is certainly quite interesting. The key point of such
research  lies in verifying existence of Nambu-Goldstone bosons, i.e.,
determining the physical masses of the fermionic scalar and pseudoscalar bound
states which appear as the products of spontaneous symmetry breaking.
In a model of NJL form, the mass determination can be simply made by using
Schwinger-Dyson equations, i.e., calculating directly the gap equation of fermion
mass and the bound state propagators induced by the four-fermion interactions.
On the other hand, it is possible that a scalar or pseudoscalar bound state is
composed of fermions with different masses or is a combination of the
scalar or pseudoscalar bilinears of these fermions [8-10]. Although a
conventional
effective potential approach is applicable to such models if one introduces
axialiary scalar fields and replaces the four-fermion interactions by Yukawa
couplings between axialiary scalar fields and  fermion fields [6],
it is unknown, by the introduction of axialiary scalar fields, whether one
could include the effect of the mass diffrence of the constituent fermions of
a  bound state.  However, we believe that the Schwinger-Dyson equation can
surely do that.  Therefore, to determine directly
the masses of the bound states and to examine fully the effect of the
fermion mass difference at finite temperature,  we prefer the Schwinger-Dyson
equation at finite temperature to a temperature effective potential.  All the
calculations will be conducted in the fermion bubble diagram approximation
which amounts to the leading order of the 1/N expansion. \\
\indent   Since the determination of physical masses of the bound states is
the key point of our research;  it is certainly more convenient to take the
real-time formalism of thermal field theory [4] than the imaginary-time
formalism. In this way we will be able to avoid the cumbersome analytic
continuation which is necessary in the latter formalism.  However,  again,
because of the possible mass differences of the constituent fermions inside the
bound states, the use of the real-time formalism will present a new question,
i.e., whether this formalism consistent is with Nambu-Goldstone mechanism.
In  a simple model with ${\rm U}_L(N)\times {\rm U}_R(N)$-invariant four-fermion
interactions, it has been proved that in the real-time formalism of thermal
field theory the Nambu-Goldstone mechanism works indeed, but it is under the
assumption that the fermions are of equal masses and  equal chemical
potentials [11]. In this paper, among other things, to examine further the
stated  consistency, we will consider a more realistic model -- a one-generation
fermion condensate scheme of electroweak symmetry breaking [9].  The extension
to the many-generation case [10] is direct. \\
\indent The paper is arranged as follows.  In Sec. II we present the model,
give its Lagrangian and derive the gap equation at  finite temperature.
In Secs. III, IV and V we will respectively calculate propagators of the scalar,
the pseudoscalar, and the charged-scalar bound states at finite temperature by
means of the Schwinger-Dyson equations satisfied by the fermionic four-point
Green functions, determine the bound states' masses, and discuss the pinch
singularity problem which is especially related to the real-time formalism of
thermal field theory.  Finally, in Sec. VI we come to our conclusions.\\
\section{One-generation fermion condensate model and gap equation}
In this model of electroweak symmetry breaking, we have one generation of
$Q$ fermions forming a ${\rm SU}_L(2)\times {\rm U}_Y(1)$  doublet $(U,D)$ and are assigned
in the representation $R$ of the color group ${\rm SU}_c(3)$ with the dimension
$d_Q(R)$.   The symmetry breaking is induced by the effective four-fermion
interactions among the $Q$ fermions below some high momentum scale $\Lambda$
which, in real-time thermal field theory, are  described by the Lagrangian
[4,9,10]
\begin{equation}
{\cal L}_{4F}= {\cal L}^S_{4F} + {\cal L}^P_{4F} + {\cal L}^C_{4F}.
\end{equation}
The neutral scalar couplings \[
{\cal L}^S_{4F}=\frac{1}{4}\sum_{a=1}^{2}\sum_{Q,Q'}
                (-1)^{a+1}g_{Q'Q}{(\bar{Q'}Q')}^{(a)}{(\bar{Q}Q)}^{(a)},
\]
\begin{equation}
               g_{Q'Q}=g^{1/2}_{Q'Q'}
                              g^{1/2}_{QQ}, \ \ Q,Q'=U,D,
\end{equation}
where $a=1$ denotes physical fields and $a=2$ ghost fields. The physical fields
and the ghost fields interact only through the propagators.  The neutral
pseudoscalar couplings
\[
{\cal L}^P_{4F}=\frac{1}{4}\sum_{a=1}^{2}\sum_{Q,Q'}
                (-1)^{a+1}g'_{Q'Q}{(\bar{Q'}i\gamma_5Q')}^{(a)}
                                        {\bar{Q}i\gamma_5Q)}^{(a)},
\]
\begin{equation}
           g'_{Q'Q}={(-1)}^{I^3_{Q'}-I^3_Q}g_{Q'Q}, \ \ Q,Q'=U,D,
\end{equation}
and $I^3_Q$ denotes the third component of the weak isospin of the
$Q$ fermions.   The charged scalar couplings
\[
{\cal L}_{4F}^C=\frac{G}{2}\sum_{a=1}^{2}
{(-1)}^{a+1}{(\bar{D}\Gamma^+U)}^{(a)}{(\bar{U}\Gamma^-D)}^{(a)},
\]
\[
\Gamma^{\pm}=\frac{1}{\sqrt{2}}[\cos\varphi -\sin\varphi \pm(\cos\varphi +
\sin\varphi )\gamma_5],
\]
\begin{equation}
G=g_{UU}+g_{DD}, \ \ \ \ \cos^2\varphi =g_{UU}/G, \ \ \ \ \sin^2\varphi =g_{DD}/G.
\end{equation}
We indicate that Lagrangian (2.1) is the real-time thermal field theory version
of the following zero-temperature four-fermion Lagrangian for $n=1$ [10]:
\begin{equation}
{\cal L}^0_{4F}=\frac{G}{4}\left[
{(\phi^0_S)}^2+{(\phi^0_P)}^2+2\phi^+\phi^-\right],
\end{equation}
where
\[
\phi^0_S=\cos\varphi {(\bar{U}U)}^{(1)}+\sin\varphi{(\bar{D}D)}^{(1)},
\]
\[
\phi^0_P=\cos\varphi {(\bar{U}i\gamma_5U)}^{(1)}-
         \sin\varphi{(\bar{D}i\gamma_5D)}^{(1)},
\]
\begin{equation}
\phi^-={(\bar{U}\Gamma^-D)}^{(1)}, \ \ \phi^+={(\bar{D}\Gamma^+U)}^{(1)}
\end{equation}
are, respectively, the configurations of the physical neutral scalar, neutral
pseudoscalar, and charged scalar bound states which are expressed by physical
fermion fields with $a=1$.
In zero-temperature field theory, one assumes that ${\cal L}^S_{4F}$
will lead to the vacuum expectation value $\sum_{Q=U,D}g_{QQ}
\langle{(\bar{Q}Q)}^{(1)}\rangle \neq 0$ and this will induce spontaneous
breaking of electroweak group.  At finite temperature $T$ and in the real-time
formalism of thermal field theory, we will assume the thermal expectation
value $\sum_{Q=U,D}g_{QQ}{\langle{(\bar{Q}Q)}^{(1)}\rangle}_{T} \neq 0$,
where only the physical fields $(a=1)$ are considered [12]. We thus obtain the
mass of the Q fermion,
\begin{equation}
m_Q(T, \mu)\equiv m_Q=-\frac{1}{2}g^{1/2}_{QQ}\sum_{Q'=U,D}g^{1/2}_{Q'Q'}
                      {\langle(\bar{Q'}Q')^{(1)}\rangle}_T ,
\end{equation}
which will lead to the relation
\begin{equation}
m_Q/g^{1/2}_{QQ}=m_{Q'}/g^{1/2}_{Q'Q'}
\end{equation}
and the gap equation at finite temperature $T$,
\begin{equation}
1=\sum_{Q=U,D}g_{QQ}I_Q,
\end{equation}
with
\begin{eqnarray}
I_Q&=&-\frac{1}{2m_Q}{\langle{(\bar{Q}Q)}^{(1)}\rangle}_T
   =\frac{d_Q(R)}{2m_Q}\int \frac{d^4l}{{(2\pi)}^4}{\rm tr}[iS^{11}(l,m_Q)]\nonumber
                 \\
  &=&2d_Q(R)\int \frac{d^4l}{{(2\pi)}^4}\left[
     \frac{i}{l^2-m_Q^2+i\varepsilon}-2\pi\delta (l^2-m^2_Q)\sin^2\theta(l^0,
     \mu_Q)\right],
\end{eqnarray}
where we have used the thermal propagator of fermion in the matrix form
\begin{eqnarray}
\left(\matrix{
iS^{11}(l,m_Q), & iS^{12}(l,m_Q) \cr
iS^{21}(l,m_Q), & iS^{22}(l,m_Q) \cr
}\right)&=&
\left(\matrix{
i/(\not\!{l}-m_Q+i\varepsilon), & 0 \cr
0, & -i/(\not\!{l}-m_Q-i\varepsilon) \cr
}\right) \nonumber \\
  &&-2\pi(\not\!{l}+m_Q)\delta (l^2-m_Q^2)\left(\matrix{
\sin^2\theta(l^0,\mu_Q), & \frac{1}{2}e^{\beta\mu_Q/2}\sin2\theta(l^0, \mu_Q) \cr
-\frac{1}{2}e^{-\beta\mu_Q/2}\sin2\theta(l^0, \mu_Q),& \sin^2\theta(l^0,\mu_Q) \cr
}\right),
\end{eqnarray}
with the chemical potential $\mu_Q$ of the $Q$ fermion and the denotations
\begin{equation}
\sin^2\theta(l^0,\mu_Q)=\frac{\theta(l^0)}{\exp[\beta (l^0-\mu_Q)]+1}
                        +\frac{\theta(-l^0)}{\exp[\beta (-l^0+\mu_Q)]+1}
\end{equation}
and $\beta=1/T$.  The gap equation (2.9) could be satisfied merely at lower
temperature $T$ than $T_c$, where $T_c$ is the critical temperature above
which Eq. (2.9) is no longer valid and thus electroweak symmetry restoration is
implied [12,13].  In view of this, in the following discussion we will assume
$T< T_c$ so that the gap equation (2.9) can always be used.
\section{Scalar bound state mode}
The propagators for fermionic bound states correspond to the four-point
Green functions of the fermions. Since there exist two types of four-fermion
interaction vertices $(a=1,2)$ in  real-time thermal field theory, the
four-point Green functions will also be a matrix with the row and the column
denoted by the index $a$.  The four-point functions for the transition
from ${(\bar{Q}Q)}^{(a)}$ to ${(\bar{Q'}Q')}^{(b)}$ can be denoted by
$\Gamma_S^{Q'bQa}(p)$; then, from Eq. (2.2), they will obey the following
linear algebraic equations [10]:
\begin{equation}
\sum_c\sum_{Q''}\Gamma_S^{Q'bQ''c}(p)\left[
\delta_{Q''Q}\delta^{ca}-N_{Q''}^{ca}(p)g_{Q''Q}{(-1)}^{a+1}\right]=
\frac{i}{2}g_{Q'Q}\delta^{ba}{(-1)}^{a+1},
  \ \   Q',Q=U,D, \ \ b,a=1,2,
\end{equation}
where $p$ is the four-momentum of the bound state, and $-2iN_Q^{ca}$ represents
the contribution of the $Q$-fermion loop with an $a$-type and a $c$-type
scalar coupling vertex [Eq.(2.2)], i.e.,
\begin{equation}
N_Q^{ca}(p)=-\frac{i}{2}d_Q(R)\int \frac{d^4l}{{(2\pi)}^4}
            tr\left[iS^{ca}(l,m_Q)iS^{ac}(l+p,m_Q)\right].
\end{equation}
Equations (3.1) have the solutions
\begin{eqnarray}
\Gamma_S^{Q'bQa}(p)&=&\frac{i}{2\Delta_S(p)}g_{Q'Q}\left\{
\left(\delta^{b1}\left[1+\sum_{Q}g_{QQ}N_Q^{22}(p)\right]
-\delta^{b2}\sum_{Q}g_{QQ}N_Q^{21}(p)\right)\delta^{a1}\right. \nonumber \\
&&-\left.\left(\delta^{b2}\left[1-\sum_{Q}g_{QQ}N_Q^{11}(p)\right]
  +\delta^{b1}\sum_{Q}g_{QQ}N_Q^{12}(p)\right)\delta^{a2}\right\},
  \ \ Q',Q=U,D, \ \ b,a =1,2,
\end{eqnarray}
where the coefficient determinant of Eqs. (3.1),
\begin{equation}
\Delta_S(p)=\left[1-\sum_{Q}g_{QQ}N_Q^{11}(p)\right]
            \left[1+\sum_{Q}g_{QQ}N_Q^{22}(p)\right]+
            \left[\sum_{Q}g_{QQ}N_Q^{12}(p)\right]
            \left[\sum_{Q}g_{QQ}N_Q^{21}(p)\right].
\end{equation}
The propagator for the physical scalar bound state $\phi_S^0$ expressed in
Eq. (2.6) is
\begin{eqnarray}
\Gamma^{\phi_S^0}(p)&=&\cos^2\varphi\Gamma_S^{U1U1}(p)+
                     \sin^2\varphi\Gamma_S^{D1D1}(p)+
                     \sin\varphi\cos\varphi\left[
                     \Gamma_S^{D1U1}(p)+\Gamma_S^{U1D1}(p)
                     \right]  \nonumber \\
       &=&iG\left[1+\sum_{Q}g_{QQ}N_Q^{22}(p)\right]/2\Delta_S(p).
\end{eqnarray}
The problem is reduced to the calculation of $N_Q^{ab}(p)$.  By using the
formula
\begin{equation}
\frac{1}{X+i\varepsilon}=\frac{X}{X^2+\varepsilon^2}-i\pi\delta(X)
\end{equation}
and through direct but rather lengthy derivation we obtain
\begin{eqnarray}
N_Q^{11}(p)&=&I_Q+\frac{1}{2}(p^2-4m^2_Q+i\varepsilon)\left[
              K_Q(p)+H_Q(p)-iS_Q(p)\right] \nonumber \\
           &=&-{\left[N_Q^{22}(p)\right]}^*, \nonumber \\
N_Q^{12}(p)&=&N_Q^{21}(p)=-\frac{i}{2}(p^2-4m_Q^2)R_Q(p), \end{eqnarray}
where
\begin{eqnarray}
K_Q(p)&=&-2d_Q(R)\int\frac{id^4l}{{(2\pi)}^4}\frac{1}{(l^2-m_Q^2+i\varepsilon)
         [(l+p)^2-m_Q^2+i\varepsilon]} \nonumber \\
      &=&\frac{d_Q(R)}{8\pi^2}\int \limits_{0}^{1}dx\left(
          \ln\frac{\Lambda^2+M_Q^2}{M_Q^2}-\frac{\Lambda^2}{\Lambda^2+M_Q^2}
          \right), \ \ M_Q^2=m_Q^2-p^2x(1-x),
\end{eqnarray}
with the four-dimension Euclidean momentum cutoff $\Lambda$,
\begin{eqnarray}
H_Q(p)&=&4\pi d_Q(R)\int \frac{d^4l}{{(2\pi)}^4}\left\{
         \frac{(l+p)^2-m_Q^2}{{[(l+p)^2-m_Q^2]}^2+\varepsilon^2}+(p\to -p)
         \right\}\delta(l^2-m_Q^2)\sin^2\theta(l^0,\mu_Q) \nonumber \\
      &=&\frac{1}{16\pi^2|\stackrel{\rightharpoonup}{p}|}
         \int \limits_{0}^{\infty}
         \frac{d|\stackrel{\rightharpoonup}{l}||\stackrel{\rightharpoonup}{l}|}
              {\omega_{Ql}}\left[\ln
         \frac{{(p^2-2\omega_{Ql}p^0+2|\stackrel{\rightharpoonup}{l}|
               |\stackrel{\rightharpoonup}{p}|)}^2+\varepsilon^2}
              {{(p^2-2\omega_{Ql}p^0+2|\stackrel{\rightharpoonup}{l}|
               |\stackrel{\rightharpoonup}{p}|)}^2+\varepsilon^2}
              +(p^0\to -p^0) \right] \nonumber \\
      &&\left[\frac{1}{\exp[\beta(\omega_{Ql}-\mu_Q)]+1}+
              \frac{1}{\exp[\beta(\omega_{Ql}+\mu_Q)]+1}
        \right], \ \ \omega_{Ql}=\sqrt{{\stackrel{\rightharpoonup}{l}}^2+m_Q^2},
\end{eqnarray}
\begin{equation}
S_Q(p)=4\pi^2d_Q(R)\int
       \frac{d^4l}{{(2\pi)}^4}\delta(l^2-m_Q^2)\delta[(l+p)^2-m_Q^2]
       [\sin^2\theta(l^0+p^0,\mu_Q)\cos^2\theta(l^0,\mu_Q)+\cos^2\theta(l^0+p^0,
       \mu_Q)\sin^2\theta(l^0, \mu_Q)],
\end{equation}
and
\begin{equation}
R_Q(p)=2\pi^2d_Q(R)\int
\frac{d^4l}{{(2\pi)}^4}\delta(l^2-m_Q^2)\delta[(l+p)^2-m_Q^2]
\sin 2\theta(l^0,\mu_Q)\sin 2\theta(l^0+p^0, \mu_Q).
\end{equation}
In Eq.(3.9) the limit $\varepsilon \to 0$ will be taken only after the total
calculations are completed.  Substituting Eq. (3.7) into Eq. (3.5) and using
the gap equation (2.9) and the relation
\begin{equation}
g_{QQ}/G=m_Q^2/\sum_{Q}m_Q^2
\end{equation}
derived from Eq. (2.8), we obtain
\begin{eqnarray}
\Gamma^{\phi_S^0}(p)&=-i\sum_{Q}m_Q^2/&\left\{
    \sum_{Q}(p^2-4m_Q^2+i\varepsilon)m_Q^2\left[K_Q(p)+H_Q(p)-iS_Q(p)\right]
    \right.  \nonumber \\
                     &&  \left.-\frac{{[\sum_{Q}(p^2-4m^2)m_Q^2R_Q(p)]}^2}
            {\sum_{Q}(p^2-4m^2-i\varepsilon)m_Q^2\left[K_Q(p)+H_Q(p)+iS_Q(p)
            \right]}
   \right\}.
\end{eqnarray}
The mass squared $m^2_{\phi_S^0}$ is determined by the pole's position of
$\Gamma^{\phi_S^0}(p)$, i.e., by the equation
\begin{equation}
{\left\{\sum_{Q}(p^2-4m_Q^2)m_Q^2[K_Q(p)+H_Q(p)]\right\}}^2+
{\left\{\sum_{Q}(p^2-4m_Q^2)m_Q^2S_Q(p)\right\}}^2=
{\left\{\sum_{Q}(p^2-4m_Q^2)m_Q^2R_Q(p)\right\}}^2.
\end{equation}
We notice that when $\Lambda^2 >> M_Q^2$,
\begin{eqnarray}
K_Q(p)&=&\frac{d_Q(R)}{8\pi^2}\left(\ln \frac{\Lambda^2}{m_Q^2}+1\right)\nonumber \\
      && -\frac{d_Q(R)}{8\pi^2}\left\{\matrix{
       \sqrt{\lambda_Q-1}\arctan \frac{1}{\sqrt{\lambda_Q-1}}
      & {\rm if}  & \lambda_Q>1,\cr
       \sqrt{1-\lambda_Q}\ln \frac{1+\sqrt{1-\lambda_Q}}{1-\sqrt{1-\lambda_Q}},
      & {\rm if}  & \lambda_Q<1, \cr}
      \right., \lambda_Q=4m_Q^2/p^2.
\end{eqnarray}
Hence $K_Q(p)$ is real and positive definite. The same conclusion is true
with $H_Q(p)$ and $S_Q(p)$ since the integrands in Eqs. (3.9) and (3.10)
are real and positive, but not applicable to $R_Q(p)$.  In addition, we have
the inequalities
\begin{equation}
S_Q(p)\pm R_Q(p)=4\pi^2d_Q(R)\int \frac{d^4l}{{(2\pi)}^4}
                \delta(l^2-m_Q^2)\delta[{(l+p)}^2-m_Q^2]
                \sin^2[\theta(l^0+p^0,\mu_Q)\pm \theta(l^0, \mu_Q)]\geq 0.
\end{equation}
Let
\[
\sum_{Q}m_Q^2K_Q(p)=k,\ \ \sum_{Q}m_Q^2H_Q(p)=h,\ \ \sum_{Q}m_Q^2S_Q(p)=s,
\ \ \sum_{Q}m_Q^2R_Q(p)=r,
\]
\begin{equation}
\sum_{Q}m_Q^4K_Q(p)=\tilde{k},\ \ \sum_{Q}m_Q^4H_Q(p)=\tilde{h},\ \
\sum_{Q}m_Q^4S_Q(p)=\tilde{s}, \ \ \sum_{Q}m_Q^4R_Q(p)=\tilde{r}.
\end{equation}
then Eq. (3.14) can be rewritten as
\begin{equation}
[(k+h)p^2-4(\tilde{k}+\tilde{h})]^2+{(sp^2-4\tilde{s})}^2=
{(rp^2-4\tilde{r})}^2.
\end{equation}
It has the solutions
\begin{eqnarray}
p^2&=&4\frac{[(k+h)(\tilde{k}+\tilde{h})+s\tilde{s}-r\tilde{r}]}
          {{(k+h)}^2+s^2-r^2}\{1\pm A\}, \nonumber \\
    &&A={\left\{1-\frac
{[(k+h)^2+s^2-r^2][(\tilde{k}+\tilde{h})^2+\tilde{s}^2-\tilde{r}^2]}
{{[(k+h)(\tilde{k}+\tilde{h})+s\tilde{s}-r\tilde{r}]}^2}
  \right\}}^{1/2}.
\end{eqnarray}
Equation (3.19) shows that we could obtain two different $m_{\phi_S^0}^2$. However,
we indicate that, first of all, in the special case where only  single-flavor
$Q$ fermions exist (e.g., in the top-quark condensate scheme [8]) or all the $m_Q$
are equal (mass degenerate), the $A$ will be identical to zero and we still
have only a single $m_{\phi_S^0}$. Next, in general case, if the momentum cut-off
$\Lambda$ is large enough, then, considering Eq. (3.16), we always have
\[
(k+h)^2 >> s^2-r^2>0, \ \ (\tilde{k}+\tilde{h})^2>>\tilde{s}^2-\tilde{r}^2>0,
\ \ (k+h)(\tilde{k}+\tilde{h})>>s\tilde{s}-r\tilde{r},
\]
and thus $A\approx 0$. In fact, $A\neq 0$ can be explained as thermal fluctuation
of the squared mass of $\phi_S^0$.  In any way, we may consider that physically
the mass of $\phi_S^0$ is determined by the equation
\begin{eqnarray}
m^2_{\phi_S^0}=p^2&\simeq &\frac{4}{{(k+h)}^2+s^2-r^2}
          \left.[(k+h)(\tilde{k}+\tilde{h})+s\tilde{s}-r\tilde{r}]\right|_{
          p^2=m^2_{\phi_S^0}} \nonumber \\
      &=&\left.
         \frac{4(k+h)(\tilde{k}+\tilde{h})+2[(s+r)(\tilde{s}-\tilde{r})+
               (\tilde{s}+\tilde{r})(s-r)]}
              {(k+h)^2+s^2-r^2}\right|_{p^2=m^2_{\phi_S^0}}.
\end{eqnarray}
Based on Eq. (3.16) we have $s-r\geq 0$ and $\tilde{s}-\tilde{r}\geq 0$; thus
it is deduced from Eq. (3.20) that
\begin{equation}
2(m_Q)_{\rm min}\leq m_{\phi_S^0} \leq 2(m_Q)_{\rm max},
\end{equation}
where $(m_Q)_{\rm min}$ and $(m_Q)_{\rm max}$ are, respectively, the minimal
and the maximal mass among the $Q$ fermions. The limitation (3.21) is
formally the same as the one at zero temperature [10], but  it should be
understood that $m_Q\equiv m_Q(T,\mu_Q)$ is now the $Q$-fermion mass at
$T\neq 0$. \\
\indent When $p\to 0$, by Eq. (3.9), $H_Q(p)=0$ and by Eqs. (3.10) and (3.11),
$S_Q(p)$ and $R_Q(p)$ contain the pinch singularities.  However, we obtain
from Eq. (3.16) that
\begin{equation}
\left[S_Q(p)-R_Q(p)\right]|_{p\to 0}=0 \ \ {\rm or} \ \ (\tilde{s}-\tilde{r})
|_{p\to 0}=0
\end{equation}
and this will make the pinch singularities contained in the denominator
of $\Gamma^{\phi_S^0}(p)$ cancel each other.  This result is just expected
in real-time thermal field theory.\\
\indent In addition, we also indicate that it is easy to verify by Eq. (3.3)
that, similar to the zero-temperature case, for the orthogonal combination
to $\phi_S^0$,
\begin{equation}
\tilde{\phi}^0_S=-\sin\varphi(\bar{U}U)^{(1)}+\cos\varphi(\bar{D}D)^{(1)},
\end{equation}
its propagator $\Gamma^{\tilde{\phi}^0_S}(p)=0$, i.e., the configuration
$\tilde{\phi}_S^0$, does not exist. We only have the single neutral scalar
bound state $\phi_S^0$ left.
\section{Pseudoscalar bound state mode}
The calculation of the propagator for pseudoscalar bound state is similar to
the one for scalar bound state.  The four-point function for transition from
${(\bar{Q}i\gamma_5Q)}^{(a)}$ to  ${(\bar{Q'}i\gamma_5Q')}^{(b)}$ is denoted
by $\Gamma_P^{Q'bQa}$; then from Eq. (2.3) they will obey the linear algebraic
equations [10]
\begin{equation}
\sum_c\sum_{Q"}\Gamma_P^{Q'bQ"c}(p)\left[
\delta_{Q"Q}\delta^{ca}-N_{Q"5}^{ca}(p)g'_{Q"Q}{(-1)}^{a+1}\right]=
\frac{i}{2}g'_{Q'Q}\delta^{ba}{(-1)}^{a+1},  \ \   Q',Q=U,D, \ \ b,a=1,2.
\end{equation}
We have used the denotation
\begin{equation}
N_{Q5}^{ca}(p)=-\frac{i}{2}d_Q(R)\int \frac{d^4l}{{(2\pi)}^4}
        {\rm tr}\left[i\gamma_5iS^{ca}(l,m_Q)i\gamma_5iS^{ac}(l+p,m_Q)\right],
\end{equation}
and $-2iN_{Q5}^{ca}(p)$ represents the contribution of the $Q$-fermion loop with
an $a$-type and a $c$-type pseudoscalar coupling vertex [Eq. (2.3)].  By
comparing Eqs. (3.1) with Eqs. (4.1), it is easy to find that the solutions of
Eqs. (4.1) can be obtained from the solutions (3.3) by the substitutions
\[
N_Q^{ca}(p) \rightarrow N_{Q5}^{ca}(p), \ \ g_{Q'Q} \rightarrow g'_{Q'Q}.
\]
Thus
\begin{eqnarray}
\Gamma_P^{Q'bQa}(p)&=&\frac{i}{2\Delta_S(p)}g'_{Q'Q}\left\{
\left(\delta^{b1}\left[1+\sum_{Q}g_{QQ}N_{Q5}^{22}(p)\right]
-\delta^{b2}\sum_{Q}g_{QQ}N_{Q5}^{21}(p)\right)\delta^{a1}\right. \nonumber \\
&&-\left.\left(\delta^{b2}\left[1-\sum_{Q}g_{QQ}N_{Q5}^{11}(p)\right]
  +\delta^{b1}\sum_{Q}g_{QQ}N_{Q5}^{12}(p)\right)\delta^{a2}\right\},
  \ \ Q',Q=U,D, \ \ b,a =1,2,
\end{eqnarray}
where
\begin{equation}
\Delta_P(p)=\left[1-\sum_{Q}g_{QQ}N_{Q5}^{11}(p)\right]
            \left[1+\sum_{Q}g_{QQ}N_{Q5}^{22}(p)\right]+
            \left[\sum_{Q}g_{QQ}N_{Q5}^{12}(p)\right]
            \left[\sum_{Q}g_{QQ}N_{Q5}^{21}(p)\right].
\end{equation}
The propagator of the physical neutral pseudoscalar bound state $\phi_P^0$
defined in Eq. (2.6) is
\begin{eqnarray}
\Gamma^{\phi_P^0}(p)&=&\cos^2\varphi\Gamma_P^{U1U1}(p)+
                     \sin^2\varphi\Gamma_P^{D1D1}(p)-
                     \sin\varphi\cos\varphi\left[
                     \Gamma_P^{D1U1}(p)+\Gamma_P^{U1D1}(p)
                     \right]  \nonumber \\
       &=&iG\left[1+\sum_{Q}g_{QQ}N_{Q5}^{22}(p)\right]/
       2\Delta_P(p).
\end{eqnarray}
For the orthogonal combination ${\tilde{\phi}}_P^0$ to $\phi_P^0$, we have
\begin{equation}
\Gamma^{{\tilde{\phi}}_P^0}(p)=0, \ \
{\tilde{\phi}}_P^0=\sin\varphi{(\bar{U}i\gamma_5U)}^{(1)}+
                   \cos\varphi{(\bar{D}i\gamma_5D)}^{(1)};
\end{equation}
hence, only the single neutral pseudoscalar bound state $\phi^0_P$ exists.
The calculations of $N_{Q5}^{ab}$ in Eq. (4.2) are similar to the ones of
$N_{Q}^{ab}$ in Eq. (3.2) and the results can be obtained by the substitutions
\begin{equation}
N_{Q5}^{ab}(p) = {\left.N_{Q}^{ab}(p)\right |}_{p^2-4m_Q^2 \to p^2}.
\end{equation}
By means of the gap equation (2.9) and the relation (3.12) we obtain from Eqs.
(4.5), (4.4), (4.7), and (3.7) that
\begin{eqnarray}
\Gamma^{\phi_P^0}(p)&=&-i\sum_{Q}m_Q^2/(p^2+i\varepsilon)\left\{
    \sum_{Q}m_Q^2\left[K_Q(p)+H_Q(p)-iS_Q(p)\right]
    -\frac{{[\sum_{Q}m_Q^2R_Q(p)]}^2}
            {\sum_{Q}m_Q^2\left[K_Q(p)+H_Q(p)+iS_Q(p)\right]}
     \right\} \nonumber \\
     &=&-i\sum_{Q}m_Q^2/(p^2+i\varepsilon)\left(k+h-is-\frac{r^2}{k+h+is}\right).
\end{eqnarray}
We notice that when $p\to 0$, $h=0$ and $s=r$ based on Eq. (3.21); so the terms
containing the pinch singularities in the denominator of Eq. (4.8) will cancel
each other and $\Gamma^{\phi^0_P}(p)$ becomes finite. The expression (4.8)
shows that $p^2=0$ is a single pole of the propagator $\Gamma^{\phi^0_P}(p)$
and thus $\phi^0_P$ is a massless neutral pseudoscalar composite particle. \\
\indent It is interesting to indicate that when $p^2=0$, $S_Q(p)=R_Q(p)=0$
in Eqs. (3.10) and (3.11) since the constraints $l^2=m_Q^2, \ (l+p)^2=m_Q^2$,
and $p^2=0$ could not be submitted simultaneously and in addition, based on
Eq. (3.9), also $H_Q(p)=0$;  so we will have
\begin{equation}
\Gamma^{\phi^0_P}(p)=-i\sum_{Q}m_Q^2/(p^2+i\varepsilon)k \ \ \ \
                     {\rm if} \ \ p^2\rightarrow 0,
\end{equation}
which has the identical form to the propagator for the neutral pseudoscalar
bound state at $T=0$ [10], except that the $Q$-fermion mass $m_Q(T=0)$ is replaced
by $m_Q(T,\mu_Q)$.  The result implies that the mass of $\phi^0_P$ is not
affected by a finite temperature completely.
\section{Charged scalar bound state mode}
For the calculation of the propagator for charged scalar bound state, a new
feature is that the fermion loop is constituted by the propagators of the
$U$ and $D$ fermions possibly with different masses.
Denote the four-point function for the transition from
${(\bar{U}\Gamma^-D)}^{(a)}$ to ${(\bar{D}\Gamma^+U)}^{(b)}$ by
$\Gamma_{\phi^-}^{ba}(p)$; then, based on Eq. (2.4), they will obey the linear
algebraic equations
\begin{equation}
\sum_{c=1,2}\Gamma_{\phi^-}^{bc}(p)[\delta^{ca}-L^{ca}(p)G{(-1)}^{a+1}]=
i\frac{G}{2}\delta^{ba}{(-1)}^{(a+1)}, \ \  b,a=1,2,
\end{equation}
where
\begin{equation}
L^{ca}(p)=-\frac{i}{2}d_Q(R)\int \frac{d^4l}{{(2\pi)}^4}
          {\rm tr}\left[\Gamma^-iS^{ca}(l,m_U)\Gamma^+iS^{ac}(l+p, m_D)\right]
\end{equation}
and $-2iL^{ca}(p)$ represents the contribution of the fermion loop composed
of $U$-fermion and $D$-fermion propagators with an $a$-type $\Gamma^+$ coupling
vertex and a $c$-type $\Gamma^-$ coupling vertex. The solutions of Eqs. (5.1)
are
\begin{equation}
\Gamma_{\phi^-}^{ba}(p)=\frac{iG}{2\Delta_C(p)}\left(\left\{
[1+GL^{22}(p)]\delta^{b1}-GL^{21}(p)\delta^{b2}\right\}\delta^{a1}-
\left\{[1-GL^{11}(p)]\delta^{b2}+GL^{12}(p)\delta^{b1}\right\}\delta^{a2}
\right),
\end{equation}
where
\begin{equation}
\Delta_C(p)=[1-GL^{11}(p)][1+GL^{22}(p)]+G^2L^{12}(p)L^{21}(p).
\end{equation}
The propagator for physical charged scalar bound state $\phi^-$ is
\begin{equation}
\Gamma^{\phi^-}(p)\equiv \Gamma_{\phi^-}^{11}(p)
=iG/2\left[1-GL^{11}(p)+G^2\frac{L^{12}(p)L^{21}(p)}{1+GL^{22}(p)}\right],
\end{equation}
and for the orthogonal combination $\tilde{\phi}^-$ to $\phi^-$
we have
\begin{equation}
\Gamma^{\tilde{\phi}^-}(p)=0, \ \ \ \ \tilde{\phi}^-=\frac{1}{\sqrt{2}}\bar{U}[\cos\varphi+\sin\varphi+
               (\cos\varphi-\sin\varphi)\gamma_5]D,
\end{equation}
since ${\cal L}^C_{4F}$ contains no configuration $\tilde{\phi}^-$. It is
noticed that $\phi^-$ and its hermitian conjugate $\phi^+$ have the same
propagator and they become the only two charged scalar bound states. Direct
calculations give
\begin{eqnarray}
L^{11}(p)&=&\frac{1}{G}\sum_{Q}g_{QQ}I_Q+\frac{(p^2+i\varepsilon)}{2}
          \left[K_{UD}(p)+H_{UD}(p)-iS_{UD}(p)\right]
          +\frac{1}{2}\left[E_{UD}(p)+i\frac{{(m_U^2-m_D^2)}^2}{m_U^2+m_D^2}
          S_{UD}(p)\right]   \nonumber \\
         &=&-\left[L^{22}(p)\right]^*, \nonumber \\
L^{12}(p)&=&-\frac{i}{2}\left[p^2-\frac{{(m_U^2-m_D^2)}^2}{m_U^2+m_D^2}
                      \right]R_{UD}(p)\exp[\beta (\mu_U-\mu_D)/2],
            \nonumber \\
L^{21}(p)&=&-\frac{i}{2}\left[p^2-\frac{{(m_U^2-m_D^2)}^2}{m_U^2+m_D^2}
                      \right]R_{UD}(p)\exp[-\beta (\mu_U-\mu_D)/2],
\end{eqnarray}
where
\begin{eqnarray}
K_{UD}(p)&=&\frac{d_Q(R)}{4\pi^2}\int \limits_{0}^{1}dx
        \frac{m_U^2(1-x)+m_D^2x}{m_U^2+m_D^2}\left[
        \ln \frac{\Lambda^2+M_{UD}^2(p)}{M_{UD}^2(p)}-
            \frac{\Lambda^2}{\Lambda^2+M_{UD}^2(p)}\right],\nonumber \\
         && M_{UD}^2(p)=m_U^2(1-x)+m_D^2x-p^2x(1-x),
\end{eqnarray}
\begin{equation}
H_{UD}(p)=4\pi d_Q(R)\int \frac{d^4l}{{(2\pi)}^4}\left\{
\frac{(l+p)^2-m_D^2}{[(l+p)^2-m_D^2]^2+\varepsilon^2}
\delta(l^2-m_U^2)\sin^2\theta(l^0, \mu_U)+
(p\rightarrow -p, m_U\leftrightarrow m_D, \mu_U\leftrightarrow \mu_D)
\right\},
\end{equation}
\begin{eqnarray}
E_{UD}(p)&=&4\pi d_Q(R)\frac{m_U^2-m_D^2}{m_U^2+m_D^2}
\int \frac{d^4l}{{(2\pi)}^4} \nonumber \\
&&\left\{
\frac{[(l+p)^2-m_U^2][(l+p)^2-m_D^2]}{[(l+p)^2-m_D^2]^2+\varepsilon^2}
\delta(l^2-m_U^2)\sin^2\theta(l^0, \mu_U)
-(p\rightarrow -p, m_U\leftrightarrow m_D, \mu_U\leftrightarrow \mu_D)
\right\},
\end{eqnarray}
\begin{eqnarray}
S_{UD}(p)&=&4\pi^2 d_Q(R)\int \frac{d^4l}{{(2\pi)}^4}
          \delta(l^2-m_U^2)\delta[(l+p)^2-m_D^2] \nonumber \\
          && \left\{\sin^2\theta(l^0, \mu_U)\cos^2\theta(l^0+p^0, \mu_D)+
          \cos^2\theta(l^0, \mu_U)\sin^2\theta(l^0+p^0, \mu_D)
          \right\},
\end{eqnarray}
\begin{equation}
R_{UD}(p)=2\pi^2 d_Q(R)\int \frac{d^4l}{{(2\pi)}^4}
          \delta(l^2-m_U^2)\delta[(l+p)^2-m_D^2]
          \sin 2\theta(l^0, \mu_U)\sin 2\theta(l^0+p^0, \mu_D).
\end{equation}
Considering the gap equation (2.6) we obtain from Eqs. (5.5) and (5.7)
the propagator for $\phi^-$:
\begin{eqnarray}
\Gamma^{\phi^-}(p)&=&-i/ \left\{
(p^2+i\varepsilon)\left[K_{UD}(p)+H_{UD}(p)-iS_{UD}(p)\right]+
E_{UD}(p)+i\bar{M}^2 S_{UD}(p)\right. \nonumber \\
&& \left.-\frac{(p^2-\bar{M}^2)^2 R_{UD}^2(p)}
     {(p^2-i\varepsilon)\left[K_{UD}(p)+H_{UD}(p)+iS_{UD}(p)\right]
      +E_{UD}(p)-i\bar{M}^2 S_{UD}(p)} \right.\}
\end{eqnarray}
with
\begin{equation}
\bar{M}^2={(m_U^2-m_D^2)}^2/(m_U^2+m_D^2).
\end{equation}
The mass of $\phi^-(\phi^+)$ will be determined by the zero point of the
denominator of $\Gamma^{\phi^-}(p)$.  An interesting question is that
under what conditions $p^2 \to 0$ is the pole of $\Gamma^{\phi^-}(p)$
so that $\phi^-$ and $\phi^+$ would become massless bound states.
Let us discuss this problem in two cases. \\
(1) $m_U=m_D=m_Q$. That is, the two fermion flavors in one generation are
mass degenerate.  In this case, we have $K_{UD}(p)=K_Q(p)$, $H_{UD}(p)=H_Q(p)$,
$S_{UD}(p)=S_Q(p)$, $R_{UD}(p)=R_Q(p)$, $E_{UD}(p)=0$, and $\bar{M}^2=0$. Thus
\begin{equation}
\Gamma^{\phi^-}(p)=-i/(p^2+i\varepsilon)\left[
K_Q(p)+H_Q(p)-iS_Q(p)-\frac{R_Q^2(p)}{K_Q(p)+H_Q(p)+iS_Q(p)}\right],
\end{equation}
which has a form similar to the propagator (4.8) of pseudoscalar bound
state except that now no sum of $Q$ with the weight $m_Q^2$ exists.
Therefore, it follows from Eq. (5.14) that $p^2=0$ is the single pole of
$\Gamma^{\phi^-}(p)$ and $\phi^-$ and $\phi^+$ will be exactly massless
charged bound states --- charged Nambu-Goldstone bosons. In addition,
similar to the case of $\Gamma^{\phi^0_P}$, when $p\to 0$ the pinch
singularities appearing in $S_Q(p)$ and $R_Q(p)$ also cancel each other.\\
(2) $m_U\neq m_D$.  In this case, we notice that no pinch singularity could
appear.  This may also be seen from the expressions (5.11) and (5.12) of
$S_{UD}(p)$ and $R_{UD}(p)$  in which $\delta(l^2-m^2_U)\delta[(l+p)^2-m_D^2]$
in the integrands  will be equal to be zero if  $p=0$ and $m_U\neq m_D$.
Therefore, we  could calculate the propagator for $\phi^-$ on the condition
that the ghost fields with $a=2$ are omitted completely and obtain
\begin{eqnarray}
\Gamma^{\phi^-}(p)&=&iG/2[1-GL^{11}(p)] \nonumber \\
                  &=&-i/\left\{
                  (p^2+i\varepsilon)\left[
                  K_{UD}(p)+H_{UD}(p)-iS_{UD}(p)\right]+
                  E_{UD}(p)+i{\bar{M}}^2S_{UD}(p) \right\}.
\end{eqnarray}
Equation (5.16) may be obtained approximately from Eq. (5.13) by assuming that
the momentum cutoff $\Lambda$ in $K_{UD}(p)$ is large enough to neglect
terms containing $R^2_{UD}(p)$.  The pole of $\Gamma^{\phi^-}(p)$ is
determined by the equation
\begin{equation}
p^2= - \frac{E_{UD}(p)+i\bar{M}^2S_{UD}(p)}
            {K_{UD}(p)+H_{UD}(p)-iS_{UD}(p)}.
\end{equation}
Equation (5.17) shows that at finite temperature it is possible that the single
pole of $\Gamma^{\phi^-}(p)$ is not at $p^2=0$ and thus the masses  of $\phi^-$
and $\phi^+$ are not equal to zeros.  However, it is seen from the
right-hand side of Eq. (5.17) that as long as the momentum cutoff $\Lambda$
is large enough, the single pole of $\Gamma^{\phi^-}(p)$ could still be
approximately at $p^2=0$.  In particular, we notice that when $p^2=0$ and
$p^0=|\stackrel{\rightharpoonup}{p}| \to 0$, both $E_{UD}(p)$ and $S_{UD}(p)$
in the numerator of the right-hand side of Eq. (5.17) will approach zeros.
This means that at a low energy scale, $\phi^-$ and $\phi^+$ could still be
considered as massless bound states and identified with  charged
Nambu-Goldstone bosons.
\section{Conclusions}
We have expounded electroweak symmetry breaking at finite temperature
in a one-generation fermion condensate scheme in the real-time formalism of
thermal field theory and in the fermion bubble approximation. It is proved by
means of direct calculations of the
propagators for bound states that, at the temperature $T$ below the symmetry 
restoration temperature $T_c$, it is always possible to obtain a massive 
neutral scalar bound state $\phi_S^0$, a massless neutral pseudoscalar bound 
state $\phi_P^0$, and two massless charged scalar bound states $\phi^-$ and
$\phi^+$ if the two flavors of the one generation of fermions are 
mass degenerate.  In this case, we can precisely identify $\phi_S^0$ with the
Higgs boson and $\phi_P^0$, $\phi^{\mp}$ with the three Nambu-Goldstone bosons 
which appear as the products of the spontaneous breaking of electroweak group 
${\rm SU}_L(2)\times {\rm U}_Y(1)\to {\rm U}_{Q}(1)$. In other words, the
Goldstone theorem is
valid rigorously in the fermion mass-degenerate case. On the other hand,  when 
the two fermion flavors have unequal masses, we have seen that the Higgs boson 
will show double masses due to the effect of "thermal flactuation"  except
one of the two flavors being massless, and the two charged scalar
bosons $\phi^{\mp}$ will also not be exactly massless. However, we find that
as long as the momentum cutoff $\Lambda$ of the zero-temperature sectors of
the fermion loops is sufficiently large or one is dealing with low energy 
scales of the bound states, then it is still possible approximately to obtain 
a single-Higgs-boson mass and almost massless $\phi^{\mp}$. In this case we
can say  that the Goldstone theorem is only valid approximately at a finite
temperature. The well-known top-quark condensate scheme [8] certainely belongs
to the latter case.  Whether the appearence of such a situation originates from
the real-time formalism itself of thermal field theory deserves to be examined
further.    Nevertheless, our discussions have shown that all possible
pinch singularities cancel each other and do not emerge from the final
expressions and this is just the result expected in a real-time thermal field
theory.  It is worth researching further if the above results based on the
Schwinger-Dyson equation  in the real-time formalism of thermal field theory
could also appear in the imaginary-time formalism or in an effective potential
approach.\\
\acknowledgments
This work was done (in part) with the support of the Abdus Salam International
Centre for Theoretical Physics, Trieste, Italy. It was also partially supported
by the National Natural Science Foundation of China and by Grant No. LWTZ-1298
of the Chinese Academy of Sciences. \\

\end{document}